\documentclass[twocolumn,showpacs,preprintnumbers,amsmath,amssymb,aps,prl]{revtex4}
\usepackage{graphicx}
\begin{document}

\title{Reversible to Irreversible Flow Transition in Periodically Driven Vortices}   
\author{N. Mangan,$^{1,2}$ C. Reichhardt,$^{1}$ and    
 C. J. Olson Reichhardt$^{1}$} 
\affiliation{
{$^1$} Theoretical Division,
Los Alamos National Laboratory, Los Alamos, New Mexico 87545 \\
{$^2$} Department of Physics, Clarkson University, Potsdam, New York 13699-5820} 

\date{\today}
\begin{abstract}
We show that periodically driven superconducting vortices in the presence of quenched 
disorder exhibit a transition from reversible 
to irreversible flow under
increasing vortex density or cycle period. 
This type of behavior    
has  
recently been observed for periodically sheared colloidal suspensions and 
we demonstrate that driven vortex systems exhibit remarkably similar behavior.
We also provide evidence that the onset of irreversible behavior is a 
dynamical phase transition. 
\end{abstract}
\maketitle

\vskip2pc
Recent experiments on periodic shearing of  
colloidal suspensions in a viscous media 
where thermal effects are negligible have shown a
transition from reversible behavior to
an irreversible motion as a function of increasing particle density \cite{Pine,Gallub}. 
The reversible and irreversible regimes were identified by 
measuring the net displacement of the colloids after each cycle of
a periodic shear.
In the reversible regime, the colloids return to their 
initial position at the end of each cycle, while in the
irreversible regime, the colloids do not return to their starting point.
A stroboscopic measurement of the displacements in the irreversible regime 
reveals that the colloids undergo an anisotropic random walk, with larger
displacements in the shear direction. 
For a given colloid density, 
there is a threshold for the transition from reversible to 
irreversible behavior as a function of the 
strain amplitude or the distance
the particles are sheared 
per cycle.
The strain threshold decreases as
the colloid concentration increases, 
indicating
that colloid-colloid interactions play an important role
in the transition. 
More recent modeling of this system has provided evidence that the 
reversible to irreversible transition is a  
nonequilibrium phase transition with power law divergences near the 
transition \cite{Chaikin}. Other experiments on dilute sheared colloidal 
suspensions also indicate the importance of the particle interactions
in producing irreversible or chaotic flow behaviors \cite{brown}.       

In this work, we consider
whether the general features of the reversible and irreversible 
behaviors observed in the sheared colloidal system can be
realized in a wider class of nonequilibrium many-particle systems.
Driven particles moving over random quenched disorder provide an
ideal class of systems for studying this 
issue \cite{Fisher}.
Physical realizations of 
such systems include 
vortices in type-II superconductors \cite{Olson,Moon}, 
driven Wigner crystals \cite{Fertig}, 
magnetic bubble arrays \cite{Westervelt}, driven pattern forming
systems \cite{Menon}, and
colloids moving over random landscapes \cite{Reichhardt}. 
In the presence of strong quenched disorder, plastic flow regimes
occur in these 
systems even in the absence of any thermal 
effects. 
In the case of vortices, it has been established that transitions from plastic
flow to smectic or elastic flow states can be induced as a function of
increasing dc driving force \cite{Olson}.
It is not known
whether there could also be a reversible to irreversible 
flow transition in the presence
of some form of periodic forcing. 
  
To address this question, we examine the specific system of 
vortices interacting with
random quenched disorder, and measure
vortex displacements after successive cycles of
a periodic drive. Vortex lattices have 
been used extensively as a general system  
in which to understand various nonequilibrium many-body effects
since numerous simulations and experiments can 
be performed readily over a wide range of parameters.
It has been shown that many of the nonequilibrium effects
observed in the vortex system also occur in 
other systems such as driven colloidal assemblies \cite{Reichhardt,Ling}; 
thus, we expect that our results will be generic to other 
driven interacting point particle systems with quenched disorder.
   
The simulation system consists of a two-dimensional sample 
of size $L \times L$
with periodic boundary conditions in the
$x$ and $y$ directions.
The sample contains $N_v$ vortices at a vortex density of $n_v=N_v/L^2$.
The vortex-vortex interaction
force is 
${\bf F}^{vv}_{i} = \sum^{N_{v}}_{i\neq j}f_{0}K_{1}(R_{ij}/\lambda){\hat {\bf R}}_{ij}$,  
where $K_{1}$ is a modified Bessel function, $R_{ij}=|{\bf R}_i-{\bf R}_j|$ 
is the distance between vortex $i$ and $j$ located at ${\bf R}_i$ and 
${\bf R}_j$, ${\hat {\bf R}}_{ij}=({\bf R}_i-{\bf R}_j)/R_{ij}$, 
$f_{0} = \phi_{0}/(2\pi\mu_{0}\lambda^3)$,
$\lambda$ is the London penetration depth, 
and $\phi = h/2e$ is the flux quantum.
Lengths are measured in units of $\lambda$ and the sample size is $L=64\lambda$.
For $R_{ij} > \lambda$ the vortex interaction falls off rapidly, so
a finite cutoff on the interaction range is placed at $R_{ij}= 6\lambda$.     
The quenched disorder is modeled as 
$N_p=100$ randomly placed parabolic pinning traps
of radius $r_{p} = 0.3\lambda$ and strength $F^{p}$,
with 
${\bf F}^p_i=\sum_k^{N_p}F^p(R_{ik}/r_p)\Theta(r_p-R_{ik}){\hat {\bf R}}_{ik}$,
where $\Theta$ is the Heaviside step function, $R_{ik}=|{\bf R}_i-{\bf R}^p_k|$,
${\hat {\bf R}}_{ik}=({\bf R}_i-{\bf R}^p_k)/R_{ij}$, and ${\bf R}_k^p$ is
the location of pin $k$.
The periodic drive is produced by the Lorentz force from a
square wave current of period $\tau$, which we 
model as
${\bf F}^{ext}(t)=F^{ext}{\rm{sgn}}(\sin(2\pi t/\tau)){\hat {\bf x}}$, 
such that a vortex under the influence of only the driving force
undergoes a displacement of 
$d = F^{ext}\tau/2\eta$ in half a drive cycle. 
In this work, except where indicated, we fix $F^{ext} = 0.8f_0$ and vary $d$ by
changing $\tau$. 
The overdamped equation of motion for a single vortex $i$ is
\begin{equation} 
\eta\frac{d {\bf R}_{i}}{dt} = {\bf F}^{vv}_i + {\bf F}^{p}_i + {\bf F}^{ext}(t),  
\end{equation} 
where $\eta=1$ is the viscous damping term.  
The vortices are 
initialized in random positions.
We use the same protocol as in Ref.~\cite{Pine} to stroboscopically 
determine the vortex locations after each drive cycle and 
extract the effective diffusivities.
The mean square vortex displacement $R^d_x$ and $R^d_y$
in the $x$ and $y$ directions 
is measured as a function of the number $n$ of drive cycles: 
\begin{equation}
R^{d}_{\alpha}(n\tau) = N_v^{-1}\sum^{N_v}_{i}
[({\bf R}_{i}(t_0+n\tau) - {\bf R}_{i}(t_0))\cdot {\hat {\bf \alpha}}]^2  
\end{equation}
where $\alpha=x,y$.
From this quantity we follow Ref.~\cite{Pine} and define 
an effective diffusivity 
$D_\alpha=R^d_\alpha/2t$.
When the behavior is reversible the vortices
return to their initial positions after each drive cycle
and $R^{d}_{x,y}$ and $D_{x,y}$ are zero.
We observe an initial transient period of finite $R^d_{x,y}$, after which
the system settles into a stationary state that is either reversible or 
irreversible.  The diffusivities are measured only after this stationary
state is reached.  A similar transient behavior was 
observed in the shearing experiments of Ref.~\cite{Pine}.  
We also measure the vortex velocity in the $x$ direction given by
$V_x=N_v^{-1}\sum_i^{N_v} {\bf v}_i \cdot {\hat {\bf x}}$, 
where ${\bf v}_i$ is the velocity of vortex $i$. 
In all of our simulations, $N_v>N_p$ 
so that we avoid any regimes where the vortices are all pinned.
For all of the parameter sets considered in this work, 
the vortices in the reversible regime are moving during the drive cycle. 

\begin{figure}
\includegraphics[width=3.5in]{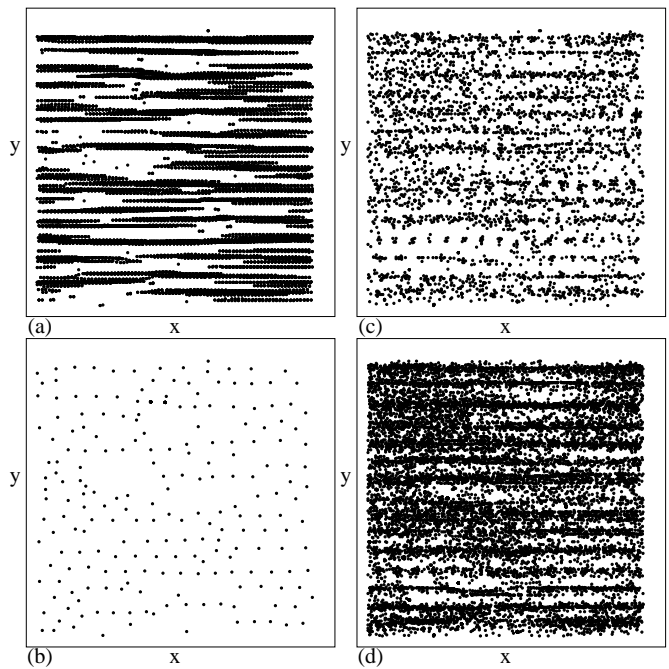}
\caption{
Black dots: Stroboscopic vortex positions marked every $0.01\tau$ in
(a) and every $\tau$ in (b,c,d).
(a) Vortex positions 
during $n=4$ cycles in 
a sample with $n_v = 0.052/\lambda^2$
and $d=48\lambda$.
(b) Vortex positions 
in the same sample as (a).  
In this reversible regime, all of the vortices return to their
initial positions after each cycle.
(c) Positions in a sample with a longer 
$d=160\lambda$
in $n=25$ cycles.  The vortices do not return to their initial
positions and the behavior is irreversible. 
(d) The same sample as (c) for $n=100$ cycles
showing that the displacements are anisotropic. 
}
\end{figure}

We find a transition from reversible to irreversible behavior as a function 
of displacement per cycle $d$ and vortex density $n_v$. 
In Fig.~1(a) we highlight the locations of the vortices every $0.01\tau$ during
$n=4$ four cycles with 
$d = 48\lambda$ and $n_v = 0.052/\lambda^2$.   
Most of the vortices are moving while a small number of vortices remain
pinned.
In Fig.~1(b) we illustrate the same system with the vortex positions
marked once per cycle during $n=100$ cycles.
Here, all of the vortices return to their initial position at the end of each cycle,
so all of the points marking the position of an individual vortex coincide.
This indicates that the system is undergoing a reversible flow.
For increasing $d$ there is a transition to an irreversible flow state. 
In Fig.~1(c) we show the stroboscopic locations of the vortices in $n=25$ cycles for the
same system as in Fig.~1(a,b) but with 
$d = 160\lambda$.  In this case, the vortices do not return to their initial
positions after each cycle.
When the number of cycles increases, the vortices continue to move further
from their
starting point after each cycle.  
Figure 1(d) illustrates the anisotropic nature of the vortex displacements
for the system in Fig.~1(c) during $n=100$ cycles. 

In Fig.~2(a) we plot the cumulative mean squared displacements $R^d_x$ and $R^d_y$ 
versus drive cycle number for the sample in the irreversible regime 
shown in Fig.~1(c,d).  
The 
dashed line 
fits
indicate that the vortices are 
performing an
anisotropic random walk 
similar to that found in the irreversible regime for the sheared colloidal system 
of Ref.~\cite{Pine}.
By conducting a series of simulations in which we vary $d$ and $n_v$,
we can extract the effective diffusivities and determine 
the thresholds for the irreversible behavior. 

\begin{figure}
\includegraphics[width=3.5in]{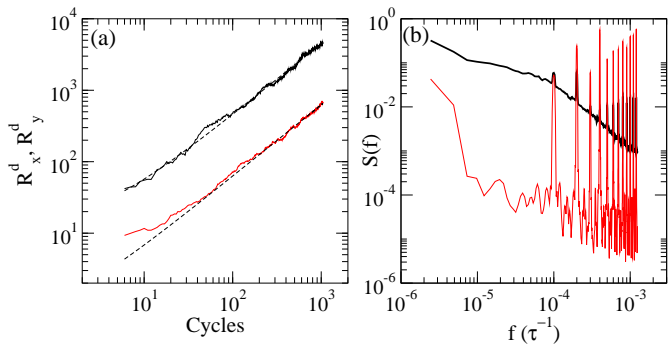}
\caption{
(a) 
$R^d_x$ (top line) and $R^d_y$ (bottom line)
versus the number of drive cycles $n$
for the system in Fig.~1(c,d) in the irreversible regime. 
The particles undergo an anisotropic random walk motion. 
Dotted lines: fits used to obtain $D_x$ and $D_y$.
(b) $S(f)$ for the irreversible regime in Fig.~1(c) (top line) and the 
reversible regime in Fig.~1(a) (bottom line).
}
\end{figure}

\begin{figure}
\includegraphics[width=3.5in]{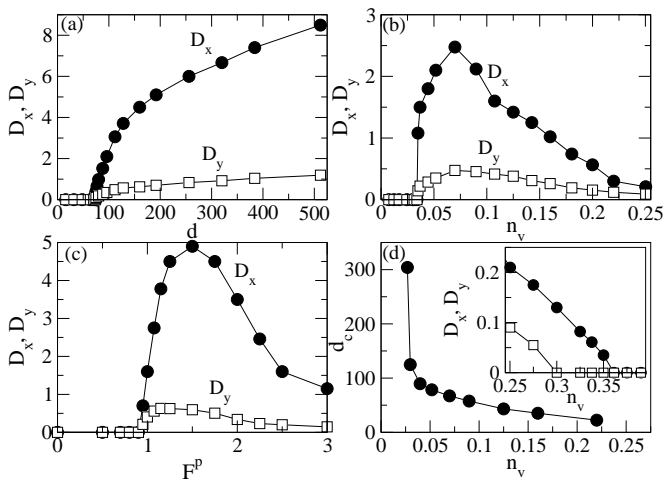}
\caption{
Diffusivities 
$D_{x}$ and $D_{y}$ versus 
$d$ 
for fixed 
$n_v = 0.052/\lambda^2$. (b) $D_x$ and $D_y$ versus
$n_v$ for fixed $d=96\lambda$. 
(c) $D_x$ and $D_y$ versus 
$F^p$ for fixed
$n_v=0.052/\lambda^2$ and $d=160\lambda$.
(d) Threshold displacements $d_{c}$ for the transition to irreversible
behavior versus $n_v$ for fixed $F^{p} = 1.25f_0$. 
Inset: $D_x$ (circles) and $D_y$ (squares) at $n_v \ge 0.25/\lambda^2$ 
continued from (b).
}
\end{figure}

In Fig.~3(a) we plot $D_x$ and $D_{y}$ versus $d$ for a sample with fixed $n_v = 0.052/\lambda^2$.
There is a clear threshold $d_c=80\lambda$ marking the transition from reversible behavior with 
$D_x=D_y=0$ to irreversible behavior with nonzero $D_x$ and $D_y$.
Above $d_c$, $D_x$ and $D_y$ monotonically increase with $d$. 
Figure 3(b) shows $D_{x,y}$ for fixed $d = 96\lambda$ and 
varied vortex density $n_v$.  
There is a threshold density $n_v=0.034/\lambda^2$ below which the system is reversible and above
which the behavior becomes irreversible.
The diffusivity passes through a peak as $n_v$ increases above the threshold, while
$D_x$ and $D_y$ decrease
at higher values of $n_v$ when the increasing strength of the vortex-vortex interactions leads to
an effective caging effect.
The inset of Fig.~3(d) illustrates the appearance of two additional phases
at higher $n_v$.  For $0.3/\lambda^2 \le n_v < 0.36/\lambda^2$, 
$D_y=0$ but $D_x$ remains
finite, resulting in a smectic phase \cite{Olson} 
where the vortices move in one-dimensional
chains that are decoupled in the $y$ direction.
For $n_v \ge 0.36/\lambda^2$, the system enters a jammed phase
where the vortex-vortex interactions 
dominate, the vortices form a rigid lattice, and the motion is reversible
again. 

In Fig.~3(c) we illustrate the effect of the pinning strength $F^p$ on the dynamics 
for a sample with $d = 96\lambda$ and $n_v = 0.052/\lambda^2$. For weak $F^{p}$ the system acts reversibly,
since when $F^{ext}>F^p$,
all the vortices depin and there is no local shear in the vortex 
lattice. 
In the colloidal shearing experiments of Ref.~\cite{Pine}, 
the shear geometry generated a velocity differential in the system,
and when the total shear was too small, the colloids moved reversibly. 
At high $F^p$, the vortices are immobilized by the pinning sites and the behavior is
reversible again.  Irreversible behavior occurs for intermediate values of $F^p$, as shown
in Fig.~3(c) by the peak in $D_x$ and $D_y$ at $F^p\approx 1.5f_0$.
In Fig.~3(d) we plot $d_c$, the threshold displacement per cycle required to induce irreversible
behavior, versus vortex density $n_v$, showing that $d_c$
decreases with increasing $n_v$. 
There is a divergence in $d_c$ near the matching density $n_v=0.0244/\lambda^2$ 
where the number of vortices equals the number of pinning sites. 
If 
$N_v < N_p$, all the vortices are pinned and the system enters a 
reversible state.
As $N_v \rightarrow N_p$ from above, 
the distance between nonpinned vortices diverges and the vortices must
be driven over longer distances in order to permit the nonpinned vortices
to interact with each other and produce the irreversible behavior.
   
In general, our results are in good agreement 
with the shear experiments of Ref.~\cite{Pine}, where a threshold
strain which decreases with increasing colloid 
density must be applied to produce irreversible motion. 
$d_{c}$ diverges as the threshold is approached from above.
Our results 
suggest that if the colloidal experiment were performed at
higher colloid density than considered in Ref.~\cite{Pine}, 
the diffusivity should decrease dramatically 
and could drop to zero if a jamming transition occurs.
It may also be possible 
that the colloidal motion will become banded, which would be analogous to the
smectic state. 

\begin{figure}
\includegraphics[width=3.5in]{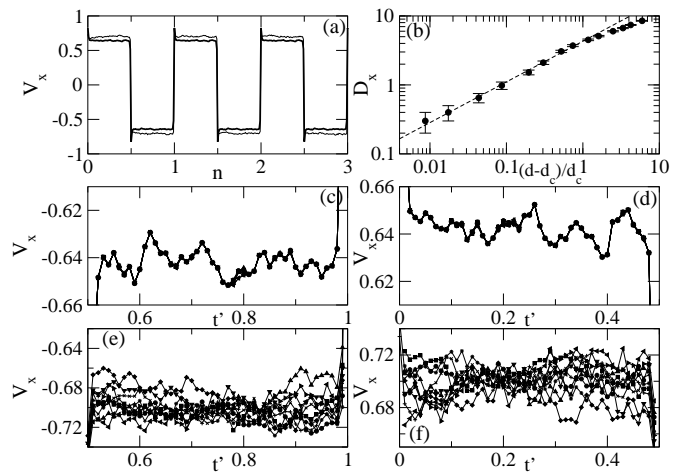}
\caption{
(a) $V_x$
versus cycle
number $n$ for (light line) the reversible regime in Fig.~1(a) and 
(heavy line) the irreversible regime in Fig.~1(c).
(b) $D_{x}$ from Fig.~3(a) vs $(d -d_{c})/d_{c}$. 
Dashed line: logarithmic fit 
$D_{x} \propto (d - d_{c})^{\beta}$ with $\beta = 0.59$. 
(c-f) Overlapping time traces of $V_x$ during 10 cycles plotted versus
$t^\prime=t-t\mod\tau$. 
Reversible regime from Fig.~1(a): (c) lower half cycle, (d) upper half cycle.
Irreversible regime from Fig.~1(c): (e) lower half cycle, (f) upper half cycle.
}
\end{figure}

In the shear experiments the transition 
from reversible to irreversible behavior can be visualized directly
by video microscopy \cite{Pine}, while for most 
vortex systems, direct imaging of the vortex motion is difficult.
A straightforward method for probing the nature of the vortex dynamics is through voltage noise 
fluctuations, where the voltage is proportional to the vortex velocity. 
In Fig.~4(a) we plot $V_x$ versus cycle number $n$ 
for the same system shown in Fig.~1(a,b) in both the
reversible and irreversible regimes. 
The velocities are sampled at a rate of $0.01\tau$.
In Fig.~4(a), we see that $V_x$ follows the square wave of the driving force and 
that the average $V_x$ for the two states is similar. 
We can superimpose the velocity curves in each cycle by plotting $V_x$ versus
$t^\prime$, where $t^\prime=t-t\mod\tau$, and compare the velocity fluctuations in the
two regimes.
In Fig.~4(c,d) we plot the superimposed fluctuations of $V_x$ for 10 cycles 
in the reversible regime for the negative and positive portions of the cycle.
Here, the fluctuations of $V_x$ are nearly identical in each cycle,
indicating that the vortices are flowing along the same trajectories during each cycle.
In the irreversible regime, shown in Fig.~4(e,f), the velocity fluctuations 
differ during each cycle and the curves do not 
overlap.
We observe a similar behavior for the fluctuations in the $y$-direction. 
Since the power spectrum of the fluctuations can be measured readily in
experiment \cite{Okuma}, we compute 
$S(f)=\left|\int |V_x(t)|e^{-2\pi i f t}dt\right|^2$ .
Fig.~2(b) shows that a broad spectrum appears in the irreversible regime,
while in the reversible regime the noise power is much lower and strong
harmonic frequencies appear.
These results show that the reversible to irreversible transition 
can be identified through noise measurements. 
A more direct visualization experiment could be performed 
by periodically driving a two-dimensional colloid system interacting
with a random substrate \cite{Ling} or optical traps \cite{Grier}.  

The colloidal shearing experiments of Refs.~\cite{Pine,Chaikin} 
found evidence of a dynamical phase transition into the irreversible state,
along with an initial transient behavior of a duration that diverges
near the transition.
We find a similar transient behavior in our system. 
If we consider $D_x$ and $D_y$  in Fig.~2(a)
as an order parameter, Fig.~4(b) shows that we find a reasonable scaling of
$D_{x} \propto (d - d_{c})^{\beta}$ with $\beta = 0.59$, 
consistent with either directed percolation in two dimensions \cite{Sana}
or conserved directed percolation.  The latter was recently proposed as
the universality class of the colloidal shear experiments \cite{Mennon}.  
It would be interesting to determine whether
a similar power law divergence occurs in the average displacements near 
the strain threshold in the colloidal system.          

In summary, we have shown with numerical simulations that a non-thermal transition 
from reversible to irreversible flow behavior occurs for periodically driven vortex
systems in the presence of quenched disorder. 
The transition is characterized 
by performing stroboscopic measurements of the vortex locations after each drive cycle. 
We find that there is a 
threshold in the displacement per cycle above which the system acts irreversibly. 
In the
irreversible regime, the vortices undergo an anisotropic random walk.  
The threshold 
decreases as the vortex density increases.
The transition can also be identified by analyzing the voltage
noise fluctuations during each cycle. 
Our results are remarkably similar to recent experimental observations of a transition
to irreversible flow for sheared colloids. In the colloid system, 
the velocity dispersion created by the shearing process permits neighboring
colloids to undergo different random displacements. 
In the vortex system, the pinning sites create velocity dispersion by generating
local shear in the vortex lattice.
The transition to the irreversible state is consistent  
with either two-dimensional directed percolation or conserved directed
percolation.
Our results suggest that the
behavior of the colloidal shear experiments may be
general to driven particle systems with quenched disorder. 
 
We thank E. Ben-Naim, M. Hastings, and D. Pine for useful discussions.
This work was carried out under the auspices of the 
NNSA of the 
U.S. DoE
at 
LANL
under Contract No.
DE-AC52-06NA25396.

\end{document}